\documentclass[11pt]{article}
 
\usepackage{times}
\usepackage{amssymb}
\usepackage{amsmath}
\usepackage{amsbsy}
\usepackage{amsfonts}
\usepackage{eucal}
\newtheorem{theo}{Theorem}
\newtheorem{lm}{Lemma}
 
\newcommand{\ket}[1]{\left\vert #1 \right\rangle}
\newcommand{\bra}[1]{\left\langle #1 \right\vert}

\newcommand{\rr}{\mathbb R}

\newcommand{\vph}{\varphi}
\newcommand{\vep}{\varepsilon}

\newcommand{\qed}{\mbox{\rule{1.6mm}{4.3mm}}}

\begin{document}

\title{Partial recovery of entanglement in bipartite entanglement 
transformations\thanks{This work was sponsored in part by the Defense 
Advanced Research Projects
Agency (DARPA) project MDA972--99--1--0017 [note that the content of this
paper does not necessarily reflect the position or the policy of the
government, and no official endorsement should be inferred], and in part
by the U.S. Army Research Office/DARPA under contract/grant number
DAAD19--00--1--0172.
\newline\mbox{\hspace{4mm}}
The research of Farrokh Vatan was performed partly at UCLA and supported 
by the above mentioned grants
and was performed partly at the Jet Propulsion Laboratory
(JPL), California Institute of Technology, under contract with National 
Aeronautics and Space Administration (NASA). The Revolutionary computing
Technologies Program of the JPL's Center for Integrated Space 
Microsystems (CISM) supported his work.
\newline\mbox{\hspace{4mm}}
Emails: som@ee.ucla.edu, vwani@ee.ucla.edu, 
        Farrokh.Vatan@jpl.nasa.gov}}
\author{Somshubhro Bandyopadhyay$^{1}$, Vwani Roychowdhury$^1$,
        Farrokh Vatan$^{1,2}$ \\
        {\small 1. Electrical Engineering Department,  UCLA, Los Angeles, CA 90095} \\
        {\small 2. Jet Propulsion Laboratory, California Institute of Technology 
                   4800 Oak Grove Drive Pasadena, CA 91109}} 
\date{ }

\maketitle

\begin{abstract}
\iffalse
We show that partial recovery of entanglement lost in {\em any} deterministic  transformation 
involving local operations and classical communication (LOCC) is {\it almost always}\/ possible.
Let $\ket{\psi}$ and $\ket{\vph}$ be $n\times n$ states and
$\ket{\psi} \longrightarrow \ket{\vph}$ under LOCC. We ask whether
there exist $k\times k$ states, $\ket{\chi}$ and $\ket{\omega}$, where $k<n$,
$E\left(\ket{\omega}\right) > E\left(\ket{\chi}\right)$ ($E$ being the entropy
of entanglement), and $\ket{\psi}\otimes\ket{\chi} \longrightarrow
\ket{\vph}\otimes\ket{\omega}$ under LOCC. 
We show that for almost all
pairs of comparable states, recovery is achievable by  $2\times 2$ states, no
matter how large the dimensions of the parent states are. For the rest of the special cases of comparable 
parent states, we show that the
 dimension of the auxiliary entangled state depends on the
presence of  equalities in the majorization relations of the parent states. We
show that recovery is still possible using states in
$k\times k$, $2<k<n$, for
 {\em all} patterns of majorization 
 relations except
only one special case.
\fi
Any  {\em deterministic} bipartite entanglement transformation 
involving finite copies of pure states and carried out using
local operations and classical communication (LOCC) results in a net loss of
entanglement. We show that for almost all
such transformations, partial recovery of lost entanglement is achievable by  
using $2\times 2$ auxiliary entangled states, no
matter how large the dimensions of the parent states are. For the rest of the special cases of 
deterministic LOCC transformations, we show that the
 dimension of the auxiliary entangled state depends on the
presence of  equalities in the majorization relations of the parent states. We
show that genuine recovery is still possible using auxiliary states in dimensions less than
that of the parent states for  {\em all} patterns of majorization 
 relations except
only one special case.
 \end{abstract}

Entanglement, shared among spatially separated parties, is 
a critical resource that enables efficient implementations of several quantum information
processing  \cite{qi} and distributed computation \cite{qc} tasks. 
To better exploit the power of entanglement, considerable effort has been put into 
understanding its transformation properties \cite{bent}--\cite{brs} and characterizing
transformations allowed under local operations and classical
communication (LOCC) . A central question is: 
what happens to the overall entanglement during transformations? In the asymptotic limit  involving 
infinite number of copies of pure states, 
entanglement can be concentrated  and diluted with unit efficiency \cite{bent}.
This remarkable asymptotic "nondissipative" property, however, does not
hold in the finite copy regime, where the process becomes inherently 
"dissipative," and a local {\em deterministic} conversion between  two pure
entangled states (which are not locally unitarily related), {\it always}
results in a net loss of entanglement \cite{nielsen}.

It is of fundamental importance to devise
local  strategies to {\it recover} the  lost entanglement in an entanglement
manipulation.  Such recovery strategies would require {\it collective}
 manipulations with ancillary resources. That is, let 
  $\ket{\psi}\sum ^{n}_{i=1}\sqrt{\alpha _{i}}\ket{i}\ket{i}$ and 
  $\ket{\vph}=\sum ^{n}_{i=1}\sqrt{\beta _{i}}\ket{i}\ket{i}$  be, respectively, 
the source and target states in $n \times n$ such that 
$\ket{\psi} \longrightarrow \ket{\vph}$ under LOCC. 
Then the amount of entanglement lost in such a transformation is
$E\left( \ket{\psi} \right) - E\left( \ket{\vph} \right)$ (where $E$ is the entropy 
of entanglement, e.g., $E\left( \ket{\psi} \right)=
-\sum_{i=1}^{n} \alpha_i \ln (\alpha_i)$), and we say that there is a {\em partial recovery} of the lost 
entanglement if there exist entangled states
$\ket{\chi}$, $\ket{\omega}$ in $k\times k$, $k < n$, such
 that  $\ket{\psi}\otimes\ket{\chi} \longrightarrow
 \ket{\vph}\otimes\ket{\omega}$
with certainty under LOCC,  and $E\left( \ket{\omega} \right) > E\left(
\ket{\chi} \right)$. Since the overall transformation involving the auxiliary states 
is dissipative, the recovered entanglement, $\left( E\left( \ket{\omega}\right) - E\left(
\ket{\chi} \right)\right)$,  is  always less than or equal to the initial amount of
lost entanglement, $\left( E\left( \ket{\psi} \right) - E\left( \ket{\vph} \right) \right)$. 
In order to minimize the use of 
ancillary resources and to reduce the complexity of the collective operations,
we consider  a partial recovery of entanglement  process
to be {\em efficient}, if the dimension of the auxiliary states, $k$, is the minimum
required for the recovery process to happen.
Moreover, in order for the partial recovery process to be {\em genuine}, we
require the 
dimension of the  auxiliary states to be smaller than that of  the parent states (i.e., 
$k<n$), since otherwise, if $k=n$ then one can  have a complete recovery of lost entanglement  by a
trivial choice:   $\ket{\chi}=\ket{\vph}$ and  $\ket{\omega}=\ket{\psi}$. 
 
A first step toward achieving partial recovery of entanglement  has recently been taken in
 \cite{morikoshi} for the {\em special case of $n=2$}.
This result is of  limited interest only:
since the auxiliary pure states are necessarily of the same dimension as the parent states
(i.e., $k=n=2$), one can always have a complete recovery of  lost entanglement
by a trivial choice of the auxiliary states. However,  
\cite{morikoshi}  presents non-trivial
selections of auxiliary states (i.e., $\ket{\chi}\neq \ket{\vph}$) that lead to partial recovery of 
entanglement.
 
We prove that genuine and efficient partial recovery of entanglement is {\em always} possible for
almost all bipartite entanglement transformations in {\em any} finite
dimension (i.e., for any $n>2$). Moreover, {\it for almost all comparable pairs}, such {\it partial
recovery} is achievable by  using auxiliary states of {\it minimum possible dimension}, i.e., $k=2$,  no
matter {\it how large} the dimensions of the parent states are. For the rest of the special cases of comparable 
parent states, we show that the  dimension of the auxiliary entangled state depends on the structure of
the majorization relations of the parent states, where the presence of
equalities in the majorization relations either in isolation or in blocks
determine the dimension of the auxiliary entanglement. 

Recall that our parent 
bipartite pure states are represented as 
 $\ket{\psi} =\sum ^{n}_{i=1}\sqrt{\alpha _{i}}\ket{i}\ket{i}$
and $\ket{\vph} =\sum ^{n}_{i=1}\sqrt{\beta _{i}}\ket{i}\ket{i},$
where $\alpha_1 \geq \alpha_2 \geq \cdots \geq \alpha_n$
and $\beta_1 \geq \beta_2 \geq \cdots \geq \beta_n$, are the
Schmidt coefficients. Hence, the eigenvalues of the reduced density matrices 
$\rho_{\psi} \equiv \mathrm{Tr}_{B(\mathrm{or}\, A)}\ket{\psi}_{AB}\bra{\psi}$
and 
$\rho_{\vph} \equiv \mathrm{Tr}_{B(\mathrm{or}\, A)}\ket{\vph}_{AB}\bra{\vph}$
are $\alpha_1,\alpha_2,\ldots,\alpha_n$ and $\beta_1,\beta_2,\ldots,\beta_n$,
respectively. Define the vector of the eigenvalues as 
$\lambda_{\psi} \equiv \left( \alpha_1,\ldots,\alpha_n\right)$
and $\lambda_{\vph} \equiv \left( \beta_1,\ldots,\beta_n\right)$. 
Since our parent states are comparable, i.e., 
$\ket{\psi} \longrightarrow \ket{\vph}$ with probability one under LOCC, it
follows from \cite{nielsen} that $\lambda_{\psi}$ is 
{\em majorized}\/ by $\lambda_{\vph}$,
(denoted as $\lambda_{\psi} \prec \lambda_{\vph}$); i.e., 
\begin{equation}
\sum ^{m}_{i=1}\alpha_i \leq \sum ^{m}_{i=1}\beta_i,
  \qquad \mbox{for every $m=1,\ldots,n-1$.}
\label{eq1}
\end{equation}
Note that  both sides equal one for  $m=n$. 

First we consider the case where $\lambda_\psi$ is  {\em strictly
majorized}\/ by $\lambda_\vph$, i.e., all the inequalities of the majorization conditions 
 (\ref{eq1}) are {\it strict}, and show that recovery with an auxiliary entangled state
in $2\times 2$ is always possible. We represent strict majorization as 
 $\lambda_\psi\sqsubset\lambda_\vph$. 
Note that {\it for a randomly picked pair of
 comparable states, the majorization inequalities are strict with probability
 one}. This guarantees that the case  where all the majorization inequalities are
strict  covers {\it almost all} possible comparable pairs. We first illustrate the basic idea involved in
the proof with a simple example.  
 
\vspace{6mm}
\noindent
{\bf Example.}
Consider the states $\ket{\psi}$ and $\ket{\vph}$ with 
$\lambda_\psi=(0.4,0.3,0.2,0.1)$, and $\lambda_\vph=(0.5,0.3,0.2,0)$. 
Then 
$\lambda_\psi \sqsubset \lambda_\vph$. Note that since $\ket{\psi} \longrightarrow
\ket{\vph}$, then for all  states  $\ket{\chi}$, $\ket{\psi}\otimes \ket{\chi}
\longrightarrow  \ket{\vph}\otimes \ket{\chi}$.  For partial recovery, our first
strategy is to obtain a $2\times 2$ state $\ket{\chi(p)}$
with $\lambda_{\chi(p)}=(p,1-p)$ and $ p \in (0.5,1)$ such that
 $\lambda_{\psi\otimes\chi(p)} \sqsubset
\lambda_{\vph\otimes\chi(p)}$. This allows a perturbation of $p$ to $p-\vep$ ($\vep>0$) 
on the right hand side of the underlying $2n$
inequalities such that  the inequalities are  still satisfied after the perturbation. 
The second requirement is  that such a perturbation should preserve
the ordering of the Schmidt coefficients,
i.e., the ordering of the coefficients is the same in both $\ket{\vph}\otimes\ket{\chi(p-\vep)}$ and 
$\ket{\vph}\otimes\ket{\chi(p)}$. These two facts together would imply that 
there is an $\vep>0$ such that 
 $\lambda_{\psi\otimes\chi(p)} \prec
\lambda_{\vph\otimes\chi(p-\vep)}$. We now let
$\ket{\chi}=\ket{\chi(p)}$ and 
 $\ket{\omega}=\ket{\chi(p-\vep)}$; since
$\vep>0$,  $E(\ket{\omega}) >
 E(\ket{\chi})$. Choosing $p = 0.8$ we have
$\lambda_{\psi\otimes\chi(p)} \sqsubset \lambda_{\vph\otimes\chi(p)}$;
 both the order of the coefficients of $\ket{\vph}\otimes\ket{\chi(0.8)}$, as well as 
the relation 
 $\lambda_{\psi\otimes\chi(p)} \prec
\lambda_{\vph\otimes\chi(p-\vep)}$
 hold if $0<\vep<0.08$. Thus, we can choose
 $\ket{\chi}=\ket{\chi(0.8)}$, and
$\ket{\omega}=\ket{\chi(0.73)}$. Then
 $\ket{\psi}\otimes\ket{\chi}
\longrightarrow \ket{\vph}\otimes\ket{\omega},$
 where $E(\ket{\omega}) >
E(\ket{\chi})$. \qed

\begin{theo}
If $\lambda_\psi$ 
is
{\em strictly majorized}\/ by $\lambda_\vph$
then there are $2\times2$ states $\ket{\chi}$ and $\ket{\omega}$ such that
$\ket{\psi}\otimes\ket{\chi}\longrightarrow\ket{\vph}\otimes\ket{\omega}$ 
and $E(\ket{\omega})>E(\ket{\chi})$.
\label{theo1}
\end{theo} 

{\bf Proof:} Let $\ket{\chi(p)}$ be a $2\times2$ state with 
$\lambda_{\chi(p)}=(p,1-p)$, and $\frac{1}{2}<p<1$. 
Note that  for all values of $p\in (\frac{1}{2}, 1)$,  
$\ket{\psi}\otimes\ket{\chi(p)}\longrightarrow 
          \ket{\vph}\otimes\ket{\chi(p)}$. The choice of $p$ determines  the orderings of the Schmidt coefficients of 
$\ket{\psi}\otimes\ket{\chi(p)}$ and $\ket{\vph}\otimes\ket{\chi(p)}$, and hence the 
inequalities in $\lambda_{\psi\otimes\chi(p)}\prec \lambda_{\vph\otimes\chi(p)}$. Conversely, one can 
think in terms of the orderings of the Schmidt coefficients. 
There is only a finite number  (in fact,  at most $n!$) of possible individual orderings of the coefficients of 
$\ket{\psi}\otimes\ket{\chi(p)}$ and $\ket{\vph}\otimes\ket{\chi(p)}$. For each such ordering
one can determine its feasible set: values of $p \in (\frac{1}{2},1)$ for which the ordering is valid.  
Each nonempty feasible set corresponds to the solution of a set of linear inequalities, and hence, is a  
union of intervals and discrete points in $(\frac{1}{2}, 1)$.  Moreover, the union of the feasible sets of 
all possible  orderings of the coefficients is the interval  $(\frac{1}{2}, 1)$.  
Hence, it follows from simple measure-theoretic arguments  that 
there exists at least one ordering, where the corresponding feasible set 
includes  intervals of nonzero lengths of the form $(a,b)$, where $\frac{1}{2} < a<b<1$. 

Next, let us restrict  $p$ to belong to  such a nontrivial  feasible set, $F$. 
We next  show  that $\lambda_{\psi\otimes\chi(p)}\sqsubset\lambda_{\vph\otimes\chi(p)}$ for
all values of  $p \in F$, {\it except} at most $2n-1$ discrete
values. Hence, the set of points $p$ where the majorization inequalities are strict
and the ordering of Schmidt coefficients is preserved is of non-zero measure, i.e., it includes intervals.
 If in the majorization relationship of 
$\lambda_{\psi\otimes\chi(p)}\prec\lambda_{\vph\otimes\chi(p)}$
one of the inequalities (among the $2n-1$ nontrivial inequalities)  is an equality, then we must
have 
\begin{equation} 
 p\sum_{j=1}^x\alpha_j+(1-p)\sum_{j=1}^y\alpha_j= 
 p\sum_{j=1}^s\beta_j+(1-p)\sum_{j=1}^t\beta_j, 
\label{equality} 
\end{equation} 
where $x+y=s+t$, $x\geq y$, and $s\geq t$.  Equivalently,
\begin{equation}
   \left(\sum_{j=1}^x\alpha_j-\sum_{j=1}^y\alpha_j-\sum_{j=1}^s\beta_j+
   \sum_{j=1}^t\beta_j\right)p = \sum_{j=1}^t\beta_j-\sum_{j=1}^y\alpha_j. 
\label{eq2}
\end{equation}
There are two cases now: {\bf (i)} Equation~(\ref{eq2}) determines a
value of $p$,  and {\bf (ii)} Equation~(\ref{eq2}) is an equivalence, and
hence, does not determine a value for $p$ . We show that case {\bf (ii)} 
is impossible: (\ref{eq2}) does not determine a value for $p$, if and only if
$ \displaystyle \sum_{j=1}^x\alpha_j = \sum_{j=1}^s\beta_j$ and 
$\displaystyle \sum_{j=1}^y\alpha_j = \sum_{j=1}^t\beta_j$. 
\iffalse
\begin{equation}
   \sum_{j=1}^x\alpha_j = \sum_{j=1}^s\beta_j \quad \mbox{and} \quad
   \sum_{j=1}^y\alpha_j = \sum_{j=1}^t\beta_j . 
\label{eq3}
\end{equation}
\fi
Since $\lambda_\psi\sqsubset\lambda_\vph$, 
%from (\ref{eq3}) 
it follows that $x>s$ and 
$y>t$. This {\em contradicts} the condition $x+y=s+t$. 
Hence, every potential equality in the majorization
relationship $\lambda_{\psi\otimes\chi(p)}\prec\lambda_{\vph\otimes\chi(p)}$ corresponds to 
a fixed value for $p$.  Since, there are at most $2n-1$ such nontrivial equalities, there are at most
$2n-1$ values for $p$ for which $\lambda_{\psi\otimes\chi(p)}\prec\lambda_{\vph\otimes\chi(p)}$
is not strict. 

Hence, there exist a $p\in F\subseteq (\frac{1}{2}, 1)$ and  an 
$0<\varepsilon<\frac{1}{2}$ such that $\lambda_{\psi\otimes\chi(p)}\prec 
                \lambda_{\vph\otimes\chi(p-\varepsilon)}$. The proof is completed by
                setting $\ket{\chi}=\ket{\chi(p)}$ and $\ket{\omega}=\ket{\chi(p-\varepsilon)}$. \qed

\vspace{6mm}
What happens if $\lambda_\psi$ 
is {\bf not} strictly majorized by $\lambda_\vph$?
We first define $\Delta_{\psi,\vph}$
as  the set of all indices $m$ such that the relation (\ref{eq1}) is an
equality:
 \[ \Delta_{\psi,\vph}=\left\{ m : 1\leq m\leq n-1,\     \sum
^{m}_{i=1}\alpha_i = \sum ^{m}_{i=1}\beta_i \right\}. \] 
Note that $1\in\Delta_{\psi,\vph}$ is
equivalent to the case $\alpha_1=\beta_1$  and $n-1\in\Delta_{\psi,\vph}$ is
equivalent  to the case $\alpha_n=\beta_n$.

We first show that {\it even in the presence of many patterns of 
equalities} in the majorization relationship of the parent states,
 {\it recovery} is {\it still possible} using only 
  {\it $2\times 2$ auxiliary states}.

\begin{theo}
Suppose that $1\not\in\Delta_{\psi,\vph}$ (i.e., $\alpha_1\neq \beta_1$), 
$n-1\not\in\Delta_{\psi,\vph}$ (i.e., $\alpha_n\neq \beta_n$), 
and if
$j\in\Delta_{\psi,\vph}$ then $j+1\not\in\Delta_{\psi,\vph}$ (i.e., there are no
consecutive equalities  in the majorization). 
Then there are 
$2\times2$ states $\ket{\chi}$ and $\ket{\omega}$ such that
$\ket{\psi}\otimes\ket{\chi}\longrightarrow\ket{\vph}\otimes\ket{\omega}$ 
and $E(\ket{\omega})>E(\ket{\chi})$.
\label{theo2}
\end{theo}

{\bf Proof:} We first show that there exists a nonempty interval $I=(\frac{1}{2}, a)$ ($1>a> \frac{1}{2}$),
such that each inequality in  the majorization relationship $\lambda_{\psi\otimes\chi(p)}\prec 
                \lambda_{\vph\otimes\chi(p)}$ is either  
                {\bf (i)} a ``{\em benign}'' identity for all $p \in I$; that is, the 
               equality holds even if on the right hand side $p$ is perturbed to $p-\vep$, for any $\vep>0$,
               or {\bf (ii)} is a {\em strict} inequality, for all $p \in I$, {\em except for}
                at most $2n-1$ discrete values. 
Such a majorization, where each inequality is either strict or a benign identity, is represented
as $\lambda_{\psi\otimes\chi(p)}\sqsubseteq \lambda_{\vph\otimes\chi(p)}$. 
One can then use the simple measure-theoretic arguments  introduced in the proof of Theorem 1, and show that
there exists an ordering of the Schmidt coefficients of   
$\ket{\vph}\otimes\ket{\chi(p)}$ such that $F\cap I$ has a nonzero measure 
(i.e., includes intervals), where $F$ is the feasible set for the given ordering. 
The two above results then show that  there exists a $p \in F\cap I$ such that 
$\lambda_{\psi\otimes\chi(p)}\sqsubseteq \lambda_{\vph\otimes\chi(p)}$, and in
a neighborhood around $p$ the ordering of the Schmidt coefficients of $\ket{\vph} \otimes \ket{\chi(p)}$
is preserved. Hence, there is an  $0<\vep<\frac{1}
{2}$, such that $\lambda_{\psi\otimes\chi(p)}\prec 
                \lambda_{\vph\otimes\chi(p-\varepsilon)}$. The proof can then be
                completed by setting $\ket{\chi}=\ket{\chi(p)}$ and $\ket{\omega}=\ket{\chi(p-\varepsilon)}$. 
We now present  a construction of such a set $I$. 

First consider the case where there are  only two equalities, i.e.,  
 $\Delta_{\psi,\vph}=\{k_1,k_2\}$, where $1<k_1<k_2<n-1$ and $k_2-k_1 \geq 2$. 
 Since $\alpha_1 \neq \beta_1$,  it cannot be the case that 
{\it both} $\alpha_1 = \alpha_{k_1}$ and $\beta_1 = \beta_{k_1}$: if it is true then
the $k_1^{th}$ inequality in the majorization is also strict and 
$k_1 \not\in \Delta_{\psi,\vph}$, which contradicts our assumption. Hence,
$\alpha_1 > \alpha_{k_1}$ or $\beta_1 > \beta_{k_1}$ or both. Similarly, 
one can argue  that {\bf (i)} since $k_1 \in \Delta_{\psi,\vph}$, and
$k_1 +1 \not\in \Delta_{\psi,\vph}$, {\it both} $\alpha_{k_1+1} = \alpha_{k_2}$ and 
$\beta_{k_1+1} = \beta_{k_2}$
cannot be true, and {\bf (ii)} since $k_2 \in \Delta_{\psi,\vph}$, and
$k_1 +1 \not\in \Delta_{\psi,\vph}$, {\it both} $\alpha_{k_2+1} = \alpha_{n}$ and 
$\beta_{k_2+1} = \beta_{n}$
cannot be true.
 Now set $I=(\frac{1}{2}, a)$, where 
\begin{multline}
a = \min \left\{ q_1 * \frac{\alpha_1}{\alpha_1+\alpha_{k_1}}, q_2* 
\frac{\beta_1}{\beta_1+\beta_{k_1}}, q_3*\frac{\alpha_{k_1+1}}{\alpha_{k_1+1}+
\alpha_{k_2}},q_4*\frac{\beta_{k_1+1}}{\beta_{k_1+1}+\beta_{k_2}}, \right. \\
\left . q_5*\frac{\alpha_{k_2+1}}{\alpha_{k_2+1}+\alpha_{n}}, 
q_6*\frac{\beta_{k_2+1}}{\beta_{k_2+1}+\beta_{n}} \right\}, 
\label{def-a-eq}
\end{multline}
and $q_i= 2$ if its accompanying multiplicative term equals $\frac{1}{2}$, otherwise $q_i=1$.  Thus, if $\alpha_1
=\alpha_{k_1}$, then $q_1=2$ and the first term, $q_1 * \frac{\alpha_1}{\alpha_1+\alpha_{k_1}}= 1$, 
plays no role in determining the value of $a$; otherwise, if 
$\alpha_1 > \alpha_{k_1}$, then $q_1=1$ and the first term, $\frac{1}{2} <
q_1 * \frac{\alpha_1}{\alpha_1+\alpha_{k_1}}< 1$, can potentially
determine $a$. {\em By construction}, $\frac{1}{2} < a <1$, and hence, $I$ is nonempty. 
The motivation of defining $a$ as above is that by restricting $p \in (\frac{1}{2}, a)$, 
it enforces a {\em partial ordering} of the Schmidt coefficients of $\ket{\vph}\otimes\ket{\chi(p)}$
and $\ket{\psi}\otimes\ket{\chi(p)}$.   For example, if $\beta_1>\beta_{k_1}$ then from (\ref{def-a-eq})
it follows that $p\beta_{k_1} < (1-p) \beta_1$, and hence,
in the ordering of the Schmidt coefficients of $\ket{\vph}\otimes\ket{\chi(p)}$, $(1-p) \beta_{1}$
will appear before $p\beta_{k_1}$. 

Next, we sketch how to show $\lambda_{\psi\otimes\chi(p)}\sqsubseteq 
\lambda_{\vph\otimes\chi(p)}$ for all $p \in F\cap I$, {\it except} at most $2n-1$ discrete
values.  If in the majorization relationship of 
$\lambda_{\psi\otimes\chi(p)}\prec\lambda_{\vph\otimes\chi(p)}$
one of the inequalities (among the $2n-1$ nontrivial inequalities)  is an equality, then 
following arguments similar to those used  in the proof of Theorem 1 and using the 
partial ordering of Schmidt coefficients enforced by the selection of $I$ (see
(\ref{def-a-eq})), one can show that either {\bf (i)} the equality determines a
value of $p$ (hence, there are at most $2n-1$ discrete values of $p$ where
any such equality can exist),  or  {\bf (ii)} the equality is a {\em benign}  
identity with one of the following forms
\begin{align}
 p\sum_{i=1}^{k_j}\alpha_i+(1-p)\sum_{i=1}^{k_j}\alpha_i &=
 p\sum_{i=1}^{k_j}\beta_i +(1-p)\sum_{i=1}^{k_j}\beta_i, \label{equ5} 
\end{align}
where $j\in \{1,2\}$. The reason  identities like (\ref{equ5}) are benign for our
purposes is that when 
$p$ is substituted by $p-\vep$ on the right hand side, then the identity still remains an equality:  
\[  p\sum_{i=1}^{k_1}\alpha_i+(1-p)\sum_{i=1}^{k_1}\alpha_i =
 (p-\vep)\sum_{i=1}^{k_1}\beta_i +(1-p+\vep)\sum_{i=1}^{k_1}\beta_i. \]
To prove the above claim, consider an equality in the majorization relationship, which as
discussed in the proof of Theorem 1 (see (\ref{eq2})), can be written as: 
\begin{equation}
   \left(\sum_{j=1}^x\alpha_j-\sum_{j=1}^y\alpha_j-\sum_{j=1}^s\beta_j+
   \sum_{j=1}^t\beta_j\right)p = \sum_{j=1}^t\beta_j-\sum_{j=1}^y\alpha_j,
\label{thm2-eq2}
\end{equation}
where $x+y=s+t$, $x\geq y$, and $s\geq t$.   Equation
(\ref{thm2-eq2}) is an equivalence if and only if the following
two conditions are simultaneously satisfied:  
(i) $\displaystyle \sum_{j=1}^t\beta_j = \sum_{j=1}^y\alpha_j$,
which is true only if $t=y \in \{0,k_1,k_2\}$,  or if $y>t$;  and 
(ii) $\displaystyle \sum_{j=1}^x\alpha_j = \sum_{j=1}^s\beta_j$,
which is true only if $x=s\in \{k_1,k_2,n\}$, or if $x>s$. 
The benign identity cases occur if  $x=y=s=t=k_j$, $j\in \{1,2\}$. Let us then
{\it consider all the other potentially feasible cases} and show that they are all
{\it impossible}: {\bf (i)} If ($y>t$ and $x \geq s$) or ($y\geq t$ and $x> s$) then we
reach the contradiction that $x+y> s+t$; {\bf (ii)} If ($y=t=0$) and ($x=s \in \{k_1,k_2,n\}$):
this implies that $(1-p)\alpha_1 < p\alpha_{k_1}$ and $(1-p)\beta_1 < p\beta_{k_1}$, which
contradicts the fact that $p\in (\frac{1}{2}, a)$ (see (\ref{def-a-eq})); {\bf (ii)} 
If ($y=t=k_1$) and ($x=s \in \{k_2,n\}$): this implies that 
$(1-p)\alpha_{k_1+1} <p \alpha_{k_2}$ {\em and} $(1-p)\beta_{k_1+1} < p\beta_{k_2}$, which
again contradicts the construction introduced in  (\ref{def-a-eq}); {\bf (iii)} 
If ($y=t=k_2$) and ($x=s=n$): this implies that 
$(1-p)\alpha_{k_2+1} < p\alpha_{n}$ {\em and} $(1-p)\beta_{k_2+1} < p \beta_{n}$, which
again contradicts the construction introduced in  (\ref{def-a-eq}).

In general, when $\Delta_{\psi,\vph}=\{k_1,k_2, \cdots, k_\ell\}$, where 
$1<k_1<\cdots< k_\ell<n-1$ and $k_{i+1}-k_i \geq 2$, then one can
show the above results for $I=(\frac{1}{2}, a)$, where 
\begin{multline*}
a = \min \left\{ q_1 * \frac{\alpha_1}{\alpha_1+\alpha_{k_1}}, q_2* 
\frac{\beta_1}{\beta_1+\beta_{k_1}},\ldots
q_{2\ell-1}*\frac{\alpha_{k_{\ell-1}+1}}{\alpha_{k_{\ell-1}+1}+\alpha_{k_\ell}},\right. \\
\left . q_{2\ell}*\frac{\beta_{k_{\ell-1}+1}}{\beta_{k_{\ell-1}+1}+\beta_{k_\ell}}, 
q_{2\ell+1}*\frac{\alpha_{k_{\ell}+1}}{\alpha_{k_{\ell}+1}+\alpha_{n}}, 
q_{2(\ell+1)}*\frac{\beta_{k_{\ell}+1}}{\beta_{k_{\ell}+1}+\beta_{n}}
 \right\},  
\end{multline*}
and $q_i$'s are chosen as in (\ref{def-a-eq}).  \qed 

\vspace{6mm}
We next show that for {\em certain equality patterns} in the majorization 
relation, {\it partial  recovery} with the help of 
{\em $2\times2$} states (or even $3\times 3$ states) 
is {\em not always possible. } 

\begin{lm}
If $\alpha_1 = \beta_1$ or $\alpha_n = \beta_n$, 
then recovery is not possible  with $2\times2$ auxiliary  
states.  Also, if both relations
$\alpha_1 = \beta_1$ and $\alpha_n = \beta_n$ hold then there is no
recovery even with $3 \times 3$ auxiliary states 
\label{lemma1}
\end{lm} 

{\bf Proof:} 
First assume that $\alpha_1 = \beta_1$ or $\alpha_n = \beta_n$. 
Suppose, by contradiction, there are 
$2\times2$ states $\ket{\chi}$ and $\ket{\omega}$ such that 
$\ket{\psi}\otimes\ket{\chi}\longrightarrow\ket{\vph}\otimes\ket{\omega}$
and $E(\ket{\omega})>E(\ket{\chi})$. 
Let $\lambda_\chi=(p,1-p)$ and $\lambda_\omega=(q,1-q)$
be the vector of eigenvectors 
of $\ket{\chi}$ and $\ket{\omega}$ with $p,q > \frac{1}{2}$. The condition 
$E(\ket{\omega})>E(\ket{\chi})$ implies that $q < p$. The relation
$\lambda_{\psi\otimes\chi} \prec \lambda_{\vph\otimes\omega}$ implies that
$\alpha_1 p \leq \beta_1 q$ and $1-\alpha_n(1-p) \leq 1-\beta_n(1-q)$.
So if $\alpha_1 = \beta_1$ or $\alpha_n = \beta_n$ then $p \leq q$ and
$E(\ket{\omega}) \leq E(\ket{\chi})$, which is a contradiction.

Now suppose that $\alpha_1 = \beta_1$ and $\alpha_n = \beta_n$.
Let us assume, by contradiction, that there are $3 \times 3$ recovery states
$\ket{\chi}$ and $\ket{\omega}$ with eigenvalue vectors
$\lambda_\chi=(p,q, 1-p-q)$ and $\lambda_\omega=(p',q',1-p'-q')$ with
$p \geq q \geq 1-p-q$ and $p' \geq q' \geq 1-p'-q'$. Then
$\lambda_{\psi\otimes\chi} \prec \lambda_{\vph\otimes\omega}$ implies that
$\alpha_1 p \leq \beta_1 p'$ and 
$1-\alpha_n(1-p-q) \leq 1-\beta_n(1-p'-q')$.
Thus $p \leq p'$ and $p+q \leq p'+q'$. Therefore, 
$\lambda_\chi \prec \lambda_\omega$ and 
$\ket{\chi} \longrightarrow \ket{\omega}$.
The last relation implies that $E(\ket{\omega}) \leq E(\ket{\chi})$ 
(see \cite{nielsen}), which is a contradiction. \qed

\vspace{6mm}
The preceding result shows that if $\alpha_1=\beta_1$ then 
any auxiliary state for partial 
recovery should be at least a $3\times3$ state. Here we show 
 that such minimum-dimension auxiliary states can exist.

\begin{theo}
If $\Delta_{\psi,\vph}=\{1\}$
then there are $3\times3$ states $\ket{\chi}$ and $\ket{\omega}$ such that
$\ket{\psi}\otimes\ket{\chi}\longrightarrow\ket{\vph}\otimes\ket{\omega}$ 
and $E(\ket{\omega})>E(\ket{\chi})$.
\label{theo3}
\end{theo}

{\bf Proof:} 
Let $\ket{\chi(p,q)}$ be a $3\times3$ state with  
$\lambda_{\chi(p,q)}=(p,q,1-p-q)$, where $p \geq q \geq 1-p-q \geq 0$.
The goal is to find the state $\ket{\omega}$ of the form 
$\ket{\chi(p,q-\vep)}$, for some $\vep > 0$. Our approach will be similar
to that introduced in the proof of Theorem 2. In particular,  it
is sufficient to show that there exists a region in the plane, $R=\{(p,q)| p \geq q \geq 1-p-q \geq 0\}$,
with nonzero area such that $\lambda_{\psi\otimes\chi(p,q)}\sqsubseteq \lambda_{\vph\otimes\chi(p,q)}$
for almost all $(p,q)\in R$, and  the set of points where it is violated
has measure zero. Thus for almost all the points in $R$, 
each inequality in  the majorization relationship 
$\lambda_{\psi\otimes\chi(p,q)}\prec 
                \lambda_{\vph\otimes\chi(p,q)}$ is either  
                {\bf (i)} a ``{\em benign}'' identity; that is, the 
               equality holds even if on the right hand side, $(p,q)$ is perturbed to $(p,q-\vep)$, for any $0<\vep$,
               or {\bf (ii)} is a {\em strict} inequality. The measure-theoretic arguments used in 
               the proofs of Theorems 1 and 2 will guarantee that there is an  $\vep >0$, 
               such that $\lambda_{\psi\otimes\chi(p,q)}\prec 
                \lambda_{\vph\otimes\chi(p,q-\varepsilon)}$. The proof can then be
                completed by setting $\ket{\chi}=\ket{\chi(p)}$ and $\ket{\omega}=\ket{\chi(p-\varepsilon)}$. 

We now present  a construction of such a set $R$. 
First note that, since $2\not\in\Delta_{\psi,\vph}$, hence $\alpha_1>\alpha_2$. Also note that
if $\alpha_2=\alpha_n$ then $\beta_2>\beta_n$. Therefore, we have to 
consider two case: {\bf (i)} $\alpha_1>\alpha_2>\alpha_n$; and 
{\bf (ii)} $\alpha_1>\alpha_2=\alpha_n$ and $\beta_2>\beta_n$. It is 
easy to check that in the case {\bf (ii)} we have $\alpha_2>\beta_n$.

 To define $R$, we choose the parameters $p$ 
and $q$ such that $p \geq q \geq 1-p-q \geq 0$ and they satisfy the 
following conditions.  
case {\bf (i)}:
\begin{equation}
 q\,\alpha_1 < p\,\alpha_2, \qquad p\,\alpha_n < (1-p-q)\alpha_2;
\label{con}
\end{equation}

%\item 
or, case {\bf (ii)}:
\begin{equation}
 q\,\alpha_1 < p\,\alpha_2, \qquad p\,\beta_n < (1-p-q)\beta_1.
\tag{\ref{con}$'$}
\end{equation}

%\end{itemize}
%
%
Note that the systems (\ref{con}) and (\ref{con}$'$) imply
$q\,\alpha_n < (1-p-q)\alpha_2$ and $q\,\beta_n < (1-p-q)\beta_2$, 
respectively. So the region $R$ is either given by 
$p \geq q \geq 1-p-q \geq 0$, and 
\begin{equation}
    q<\frac{\alpha_2}{\alpha_1}p, \qquad 
    q < 1-\frac{\alpha_2+\alpha_n}{\alpha_2}p,
\label{domain}
\end{equation}
or by
\begin{equation}
 q<\frac{\alpha_2}{\alpha_1}p, \qquad 
    q < 1-\frac{\beta_1+\beta_n}{\beta_1}p.
\tag{\ref{domain}$'$}
\end{equation}
In both cases (i.e.,  (\ref{domain}) or (\ref{domain}$'$)), $R$ defines a non--empty 
triangular region in the $(p,q)$--plane (see Figure 1).

Now for any $(p,q)\in R$, one can show that 
if any of the $3n$ inequalities in the majorization relationship
$\lambda_{\psi\otimes\chi(p,q)} \prec \lambda_{\vph\otimes\chi(p,q)}$
is an equality, then it belongs to one of the two following cases.
Case{\bf (i)}: It is a 
{\em non identical} equality, i.e., the set of $(p,q)$ that satisfies it
defines a line in $(p,q)$ plane, and hence comprises a measure zero
set. Since, the total number of such non identical equalities 
is at most $3n$, the set of all points in $R$ where there might be a non-identical
equality is of measure zero. Case{\bf {(ii)}}: It is a {\em benign} identity
of the form : 
\[ p+q\,\alpha_1+(1-p-q)\alpha_1 = p
            +q\,\beta_1+(1-p-q)\beta_1. \]
Therefore, $R$ satisfies the property that 
$\lambda_{\psi\otimes\chi(p,q)}\sqsubseteq \lambda_{\vph\otimes\chi(p,q)}$
for almost all $(p,q)\in R$. \qed

\vspace{6mm}
Let $\eta_{\psi,\vph}$ be the size of the longest block of 
consecutive equalities in the majorization relationship of $\ket{\psi}$
and $\ket{\vph}$; i.e.,  $\eta_{\psi,\vph}$ is 
the largest
integer $m$ such that  $\{j,j+1,\ldots,j+m-1\}\subseteq \Delta_{\psi,\vph}$.
Our final theorem is a general construction  which shows that if 
$\alpha_n\neq\beta_n$, then partial recovery of entanglement is always 
possible using auxiliary states of dimension $k=\eta_{\psi,\vph}+2$.
Thus, if  $\alpha_n\neq\beta_n$, then {\em 
genuine} partial recovery is {\em always possible}: since $\alpha_n\neq\beta_n$,
$\eta_{\psi,\vph}\leq (n-3)$ and {\bf hence $k<n$}. 

\begin{theo}
If $n-1\not\in\Delta_{\psi,\vph}$,
then there are $k \times k$ states $\ket{\chi}$ and $\ket{\omega}$ such that 
$k=\eta_{\psi.\vph}+2$, 
$\ket{\psi}\otimes\ket{\chi}\longrightarrow\ket{\vph}\otimes\ket{\omega}$, 
and $E(\ket{\omega})>E(\ket{\chi})$.
\label{theo4}
\end{theo}

{\bf Proof:}  We provide the proof for 
a special case. The proof can be easily extended to cover any general case .

Suppose that $\Delta_{\psi,\vph}=\{2,3,5\}$. So $\eta_{\psi,\vph}=2$.
Let $\ket{\chi(p_1,p_2,p_3)}$ be a $4 \times 4$ state with
$\lambda_{\chi(p_1,p_2,p_3)}=(p_1,p_2,p_3,p_4)$,  where $p_4=1-p_1-p_2-p_3$
and $p_1 \geq p_2 \geq p_3 \geq p_4 \geq 0$. Let
\[ S=\left\{ (p_1,p_2,p_3)\in\rr^3 : p_1\geq p_2\geq p_3\geq 1-p_1-p_2-p_3\geq0
      \right\} .\]
Then $S$ is the subset of $\rr^3$ bounded by the following polyhedral:
\begin{equation}
  \left. \begin{array}{r}
   p_1 \geq 0, \quad p_2 \geq 0, \quad p_3 \geq 0, \\ 
   \quad p_1 \geq p_2, \quad p_2 \geq p_3, \\
   \quad p_1+p_2+p_3 \leq 1, \\ p_1+p_2+2p_3 \geq 1. 
  \end{array}\right\}
\label{region}
\end{equation}
The goal is to find the 
state $\ket{\omega}$ of the form $\ket{\chi(p_1,p_2,p_3-\vep)}$, for some 
$\vep > 0$. As in the proof of Theorem 3, it suffices to show that there is a 
region $R\subseteq S$ such that 
$\lambda_{\psi\otimes\chi(p_1,p_2,p_3)}
    \sqsubseteq \lambda_{\vph\otimes\chi(p_1,p_2,p_3)}$
for almost all $(p_1,p_2,p_3)\in R$, and  the set of points where it is violated
has measure zero. The following conditions guarantee the only identical equalities 
among majorization relations of 
$\lambda_{\psi\otimes\chi(p_1,p_2,p_3)}\prec\lambda_{\vph\otimes\chi(p_1,p_2,p_3)}$ 
are benign.
\begin{equation}
 \left. \begin{array}{c}
 \frac{p_3}{p_1} < \min\left( \frac{\alpha_4}{\alpha_2}, \frac{\beta_4}{\beta_2} \right) \\
 \frac{p_4}{p_1} > \min\left( \frac{\alpha_5}{\alpha_2}, \frac{\beta_5}{\beta_2} \right) \\
 \\
 \frac{p_3}{p_2} < \min\left( \frac{\alpha_4}{\alpha_3}, \frac{\beta_4}{\beta_3} \right) \\
 \frac{p_4}{p_1} > \min\left( \frac{\alpha_5}{\alpha_3}, \frac{\beta_5}{\beta_3} \right) \\
 \\
 \frac{p_3}{p_1} < \min\left( \frac{\alpha_5}{\alpha_6}, \frac{\beta_5}{\beta_6} \right) \\
 \frac{p_4}{p_1} > \min\left( \frac{\alpha_n}{\alpha_6}, \frac{\beta_n}{\beta_6} \right) \\
 \\
 \frac{p_4}{p_3} > \min\left( \frac{\alpha_n}{\alpha_6}, \frac{\beta_n}{\beta_6} \right) \\
\end{array}\right\}
\label{conditions}
\end{equation}
The assumption $\Delta_{\psi,\vph}=\{2,3,5\}$ implies that among each pair
of numbers in the right--hand side of the above inequalities at least one 
is less than 1. Also note that for any point $(p_1,p_2,p_3)\in S$, we have
\[  \frac{p_3}{p_1}\leq\frac{p_3}{p_2} \quad \mbox{and} \quad
    \frac{p_4}{p_1}\leq\frac{p_4}{p_3} . \]
So there are real numbers $s$ and $t$ such that $0<s,t<1$ and if
\begin{eqnarray}
 \frac{p_3}{p_2} < t, \label{sol1} \\
 \frac{p_4}{p_1} > s, \label{sol2}
\end{eqnarray}
then all conditions (\ref{conditions}) are satisfied. Now we show that the conditions
(\ref{sol1}) and (\ref{sol2}) define a subset of $S$ with nonzero measure.
The condition (\ref{sol2}) defines a half--space $(1+s)p_1+p_2+p_3<1$ which 
is a subset of the half--space $\quad p_1+p_2+p_3 \leq 1$. So in the first step, 
instead of $S$, we consider the region
\begin{equation}
  \left. \begin{array}{r}
   p_1 \geq 0, \quad p_2 \geq 0, \quad p_3 \geq 0, \\
   \quad p_1 \geq p_2, \quad p_2 \geq p_3, \\
   \quad (1+s)p_1+p_2+p_3<1, \\ p_1+p_2+2p_3 \geq 1. 
  \end{array}\right\}
\label{region2}  
\end{equation}
The condition (\ref{sol1}) defines a half--space $p_2>\frac{1}{t}\,p_3$ 
which is a subset of the half--space $p_2 \geq p_3$ and cuts a non--empty 
subset $R$ of the region defined by (\ref{region}).

Like the proof of Theorem~\ref{theo2}, for some feasible set $F$ of a given
ordering of the Schmidt coefficients of 
$\ket{\phi}\otimes\ket{\chi(p_1,p_2,p_3)}$ the intersection $F\cap R$
has nonzero measure. The points $(p_1,p_2,p_3)\in F \cap R$ for which
there are nonidentical identities among majorization relations of 
$\lambda_{\psi\otimes\chi(p_1,p_2,p_3)}\prec\lambda_{\vph\otimes\chi(p_1,p_2,p_3)}$ 
belong to a {\em finite} set of hyper--planes; i.e., a measure zero set.
Also the conditions (\ref{conditions}) implies that all the other identities
among coefficients are benign.
So there is an $\vep>0$ such that
$\lambda_{\psi\otimes\chi(p_1,p_2,p_3)}\prec\lambda_{\vph\otimes\chi(p_1,p_2,p_3-\vep)}$.
\qed

\vspace{6mm}
{\bf In summary,} we have shown the following results: 

%\begin{enumerate}

%\item
1. If all the inequalities of the majorization conditions 
 (\ref{eq1}) are strict then recovery with an auxiliary entangled state
in $2\times 2$ is always possible. Note that {\it for a randomly picked pair of
 comparable states the majorization inequalities are strict with probability
 one}. 

%\item
2. If $\alpha_1 \neq \beta_1$ and $\alpha_n \neq \beta_n$
 and in the majorization conditions there are no two consecutive
 equalities, then recovery with an auxiliary entangled state in 
$2\times 2$ is possible.

%\item

3. If $\alpha_1=\beta_1$ or $\alpha_n=\beta_n$ then recovery
with a $2\times 2$ auxiliary state is not possible; and 
if $\alpha_1=\beta_1$ {\em and}\/ $\alpha_n=\beta_n$ then recovery
with a $3 \times 3$ auxiliary state is not possible.

%\item
4. If $\alpha_n \neq \beta_n$ and the number of consecutive equalities in the
majorization conditions (\ref{eq1}) are $m$, then recovery is always 
possible using  $(m+2) \times (m+2)$ auxiliary states. Hence
if $\alpha_n\neq\beta_n$, in the worst case, a recovery  by 
$(n-1)\times(n-1)$ states is achievable; i.e., a genuine recovery is
always possible.

%\end{enumerate}

Thus, we have shown a nontrivial recovery is always possible except for the
special case where $\alpha_n = \beta_n$; whether recovery is still possible
for this special case is left as an open problem. There  are many other
open questions that might be of interest. For example,  for a given pair of comparable states,
one may ask what is the maximum entanglement that can be recovered.  Similarly,
can one recover more entanglement by increasing the dimension of the auxiliary
entangled states? For example, we show that for almost all comparable states,
$2\times 2$ auxiliary states are sufficient to implement partial recovery; however,
can one have more recovery of entanglement if the dimension of the auxiliary
state is increased? We hope
that the results of the present work will lead to a better understanding of the 
subtleties involved in local  entanglement manipulation in higher dimensions.

\end{document}